\documentclass[pra,twocolumn,floatfix,a4paper,superscriptaddress]{revtex4}
\usepackage{bm,color,graphicx,amsmath,txfonts}

\newcommand{\VV}{{\cal V}}

\begin{document}

\title{Entangling two magnon modes via magnetostrictive interaction}

\author{Jie Li}
\author{Shi-Yao Zhu}
\affiliation{Zhejiang Province Key Laboratory of Quantum Technology and Device, Department of Physics and State Key Laboratory of Modern Optical Instrumentation, Zhejiang University, Hangzhou, Zhejiang, China}

\begin{abstract}
We present a scheme to entangle two magnon modes in a cavity magnomechanical system. The two magnon modes are embodied by collective motions of a large number of spins in two macroscopic ferrimagnets, and couple to a single microwave cavity mode via magnetic dipole interaction. We show that by activating the nonlinear magnetostrictive interaction in one ferrimagnet, realized by driving the magnon mode with a strong red-detuned microwave field, the two magnon modes can be prepared in an entangled state. The entanglement is achieved by exploiting the nonlinear magnon-phonon coupling and the linear magnon-cavity coupling, and is in the steady state and robust against temperature. The entangled magnon modes in two massive ferrimagnets represent genuinely macroscopic quantum states, and may find applications in the study of macroscopic quantum mechanics and quantum information processing based on magnonics.
\end{abstract}

\date{\today}
\maketitle

\section{Introduction}

Ferrimagnetic systems, for example yttrium iron garnet (YIG), provide a unique platform for the study of the strong interactions between light and matter. Owing to their high spin density (several orders of magnitude larger than those of previous spin ensembles) and low dissipation rate, in recent years the strong~\cite{Strong1,Strong2,Strong3,Strong4,Strong5,Strong6} and ultrastrong~\cite{Tobar2,Tobar3} coupling between the Kittel mode~\cite{Kittel} in the YIG sphere and the microwave cavity photons have been realized leading to cavity-magnon polaritons. This strong coupling offers a possibility to enable coherent information transfer between drastically different information carriers, and thus may find potential applications in quantum information processing, especially when the system becomes hybrid~\cite{NakaReview}, such as by coupling magnons to a superconducting qubit~\cite{Naka15,Naka17}, to phonons~\cite{Tang16,JiePRL}, or to both microwave and optical photons~\cite{Tang16b}. Furthermore, various interesting phenomena have been explored in the system of cavity-magnon polaritons, such as the observation of magnon gradient memory~\cite{TangNC}, the exceptional point~\cite{YouNC,YOU19}, manipulation of distant spin currents~\cite{spinCur}, and bistability~\cite{You18}, to name but a few.

Cavity-magnon systems of YIG spheres provide also a promising and completely new platform for the study of macroscopic quantum states~\cite{JiePRL,JieRapid}. A magnon mode can get squeezed by driving the cavity with a squeezed vacuum microwave field, and the squeezing can further be transferred to the mechanical mode if the magnetostrictive interaction is activated by driving the magnon mode with a red-detuned microwave field~\cite{JieRapid}. Such a hybrid cavity magnomechanical system can also be prepared in a genuinely tripartite entangled state by suitably driving the YIG sphere and essentially utilizing the nonlinear magnetostrictive interaction~\cite{JiePRL}. Both the entangled and squeezed states of the magnon and mechanical modes are macroscopic quantum states due to a large size of the YIG sphere and an extremely large number of spins contained (more than $10^{16}$ for a 250-$\mu$m-diam YIG sphere that has been employed in Refs.~\cite{JiePRL,JieRapid}). Alternatively, nonclassical magnon states can be prepared by coupling magnons to a superconducting qubit which provides necessary nonlinearity~\cite{Naka15,Naka17}. We note that the observation of quantum effects in massive systems has achieved significant progress in the field of cavity optomechanics~\cite{OMRMP}, where quantum squeezing of mechanical motion~\cite{sqz}, nonclassical correlations between single photons and phonons~\cite{Simon16,SimonBell}, and quantum entanglement between mechanics and a cavity field~\cite{enOM}, as well as between two massive mechanical oscillators~\cite{enMM1,enMM2} have been observed. As an analogue, cavity magnomechanics holds a potential for realizing quantum states in more massive objects. Apart from quantum effect, magnomechanically induced transparency~\cite{Tang16} and slow light effect~\cite{Wu} have been explored in such a system.

Here we present a scheme to prepare two magnon modes in two massive YIG spheres in an entangled state. To date, the proposals of preparing entangled magnon modes in cavity-magnon systems are still missing~\cite{NNNote}. The two magnon modes couple to a single microwave cavity mode via linear beamsplitter interactions, and it is known that such interactions will not yield any entanglement between the two magnon modes. Nevertheless, by activating the magnetostrictive (radiation pressure-like) interaction in one YIG sphere, we show that such an interaction can be utilized to generate entanglement between two magnon modes if one of them is suitably driven by a microwave field. The entanglement arises from the magnon-phonon coupling and partially transfers to cavity-magnon and cavity-phonon subsystems. Further, by using the effective state-swap interaction between the cavity and the other magnon mode, the two magnon modes therefore get entangled. The entanglement is in the stationary state and robust against environmental temperature. Differently from our previous work~\cite{JiePRL}, where three modes of different natures get entangled, here we aim to entangle two magnon modes in two YIG spheres.

The remainder of this paper is organized as follows. In section~\ref{MMM}, we introduce a cavity magnomechanical system of two YIG spheres, and provide its Hamiltonian and the corresponding quantum Langevin equations (QLEs). In section~\ref{results}, we study the system dynamics by adopting the linearization treatment, show how to obtain the steady-state solutions of the system, and present the main results of the entanglement between two magnon modes. In section~\ref{Kerr}, we analyse the Kerr nonlinear effect due to the strong drive of one magnon mode, and provide possible strategies to measure the entanglement. Finally, we reserve section~\ref{conc} for conclusions.

\begin{figure}[t]
\hskip-0.08cm\includegraphics[width=\linewidth]{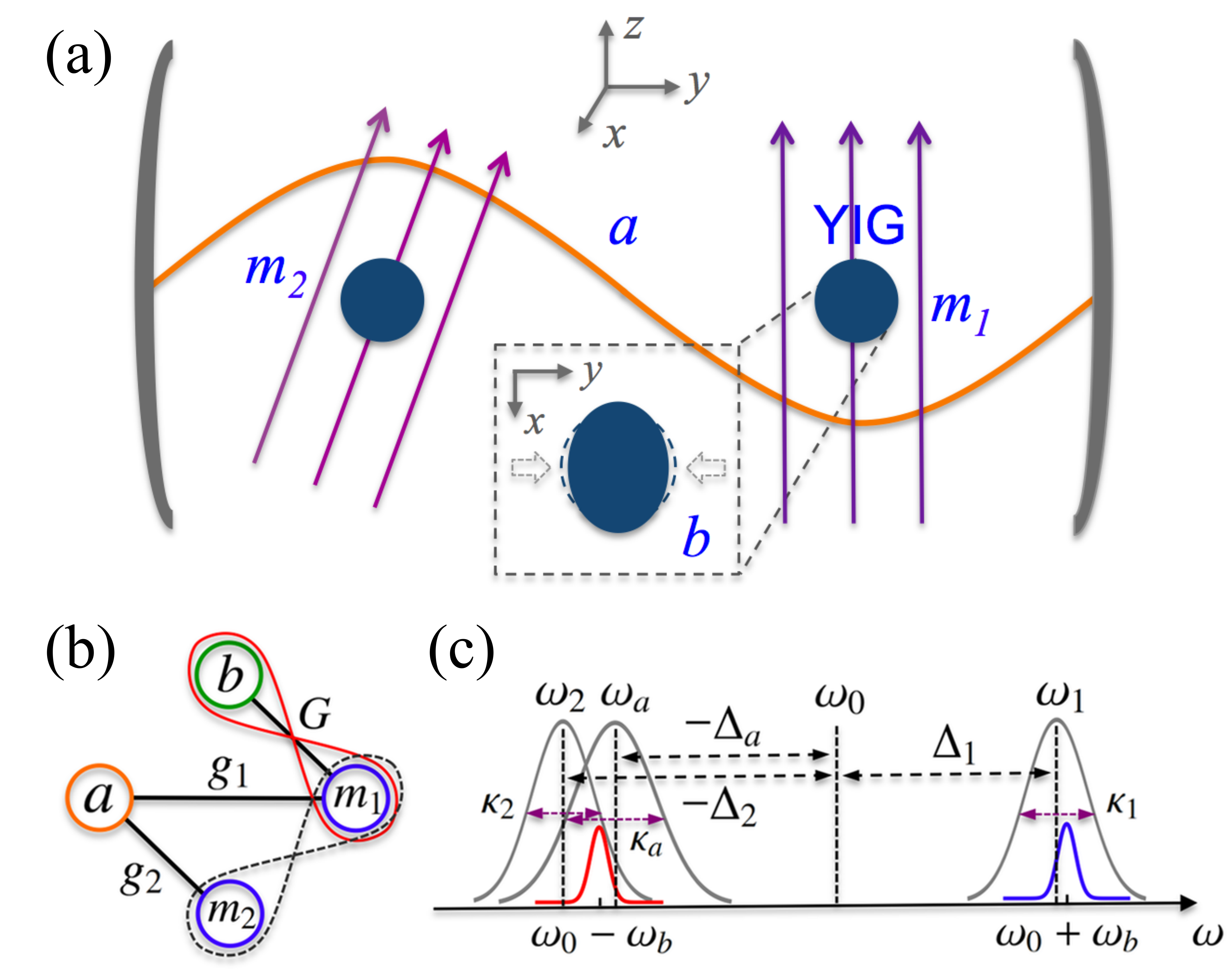} 
\caption{(a) Sketch of the system. Two YIG spheres are, respectively, placed inside a microwave cavity near the maximum magnetic fields of the cavity mode, and simultaneously in uniform bias magnetic fields, which excite the magnon modes in the spheres and couple them to the cavity mode. The directions of the bias magnetic fields are adjusted such that only in one sphere, say the right sphere, the magnetostrictive (magnon-phonon) interaction is activated, and this coupling can be enhanced by directly driving the magnon mode with a microwave source (not shown). The bias magnetic field ($z$ direction), the drive magnetic field ($y$ direction) and the magnetic field ($x$ direction) of the cavity mode are mutually perpendicular at the site of the right sphere. (b) Interactions among the subsystems. The cavity mode $a$ linearly couples to the two magnon modes $m_1$ and $m_2$ with coupling constants $g_1$ and $g_2$, respectively. Besides, the magnon mode $m_1$ couples to the mechanical mode $b$ via the nonlinear magnetostrictive interaction with an effective coupling rate $G$. The magnetostrictive interaction generates magnomechanical entanglement~\cite{JiePRL} which can be used to entangle two magnon modes. (c) Frequencies and linewidths of the system. The magnon mode $m_1$ with frequency $\omega_1$ is driven by a strong microwave field at frequency $\omega_0$, and the mechanical motion of frequency $\omega_b$ scatters photons onto the two sidebands at $\omega_0 \pm \omega_b$. If the magnon mode $m_1$ is resonant with the blue (anti-Stokes) sideband, and both the cavity with frequency $\omega_a$ and the magnon mode $m_2$ with frequency $\omega_2$ are resonant with the red (Stokes) sideband, the two magnon modes are prepared in an entangled state. }
\label{fig1}
\end{figure}

\section{The model}\label{MMM}

We consider a hybrid cavity magnomechanical system which consists of a microwave cavity mode, two magnon modes, and a mechanical mode, as shown in Fig.~\ref{fig1} (a). The two magnon modes are embodied by collective motions of a large number of spins in two macroscopic YIG spheres, and simultaneously couple to a single microwave cavity. Such a system of two YIG spheres (without involving the mechanical mode) has been used to study magnon dark modes~\cite{TangNC} and high-order exceptional points~\cite{YOU19}. The magnetic dipole interaction mediates the coupling between magnons and cavity photons [see Fig.~\ref{fig1} (b)], and this coupling can be very strong~\cite{Strong1,Strong2,Strong3,Strong4,Strong5,Strong6,Tobar2,Tobar3}. The mechanical mode is represented by the vibrations of the YIG sphere caused by the magnetostrictive force, which leads to the deformation of the geometry structure of the sphere and establishes the magnon-phonon coupling. In general, the magnetostrictive interaction is of different types depending on the resonance frequencies of the magnon and phonon modes~\cite{Kittel2}, but the {\it dispersive} magnon-phonon interaction becomes dominant when the mechanical frequency is much smaller than the magnon frequency~\cite{Tang16}, which is the case to be considered in the present work. This magnon-phonon coupling is currently small~\cite{Tang16}, but it can be efficiently enhanced by driving the sphere with a strong microwave field~\cite{JiePRL,You18}. The magnomechanical coupling strength is sensitive to the direction of the bias magnetic field~\cite{Tang16}, and we adjust the directions of the two bias magnetic fields [see Fig.~\ref{fig1} (a)] such that only in one sphere the magnetostrictive interaction is effectively activated~\cite{Note1}. The Hamiltonian of the system reads
\begin{equation}\label{Hamil}
\begin{split}
{\cal H}/\hbar &= \omega_a a^{\dag} a + \sum_{j=1,2} \omega_j m_j^{\dag} m_j + \frac{\omega_b}{2} (q^2 + p^2) + G_0 m_1^{\dag} m_1 q   \\
&+ \sum_{j=1,2} g_j (a\, m_j^{\dag} + a^{\dag} m_j) + i \Omega (m_1^{\dag} e^{-i \omega_0 t}  - m_1 e^{i \omega_0 t} ),
\end{split}
\end{equation}
where $a$ and $a^{\dag}$ ($m_j$ and $m_j^{\dag}$), $[O, O^{\dag}]\,{=}\,1$, $O\,{=}\,a$ $(m_j)$, are, respectively, the annihilation and creation operators of the cavity mode (magnon modes), $q$ and $p$, $[q, p]\,{=}\,i$, are the dimensionless position and momentum quadratures of the mechanical mode, and $ \omega_a$, $ \omega_j$, and $ \omega_b$ are the resonance frequencies of the cavity, magnon, and mechanical modes, respectively. The magnon frequencies are determined by the bias magnetic fields $H_j$ via $\omega_j \,{=}\, \gamma H_j$, where $\gamma/2\pi \,{=}\, 28$ GHz/T is the gyromagnetic ratio. The coupling rate $g_j$ denotes the linear coupling between the cavity and the $j$th magnon mode, and $G_0$ represents the single-magnon magnomechanical coupling rate. The Rabi frequency $\Omega \,\,{=}\,  \frac{\sqrt{5}}{4} \gamma \! \sqrt{N} B_0$~\cite{JiePRL} denotes the coupling strength of the drive magnetic field (with amplitude $B_0$ and frequency $\omega_0$) with the first magnon mode, where the total number of spins $N\,{=}\,\rho V$ with $\rho=4.22 \times 10^{27}$ m$^{-3}$ the spin density of the YIG and $V$ the volume of the sphere. Note that the boson (oscillator) operators $m_j$ and $m_j^{\dag}$ describe the collective motion of the spins via the Holstein-Primakoff transformation~\cite{HPT} under the assumption of the low-lying excitations $\langle m_j^{\dag} m_j \rangle \,{\ll} \, 2Ns$ (without losing generality we assume the two spheres are of the same size), where $s=\frac{5}{2}$ is the spin number of the ground state Fe$^{3+}$ ion in YIG.

From the system Hamiltonian~\eqref{Hamil}, one would see the mechanism of entanglement generation between the two magnon modes: the radiation pressure-like interaction $H_{m_1 b}=\hbar G_0 m_1^{\dag} m_1 q$ allows one to create entanglement between the magnon mode $m_1$ and the mechanical mode by suitably driving the magnon mode, similarly to creating optomechanical entanglement~\cite{DV07}. Such an interaction can also lead to cavity-magnon ($m_1$) and cavity-phonon entanglement via the linear coupling $H_{a m_1} \,{=}\,\,\hbar g_1 (a\, m_1^{\dag} + a^{\dag} m_1)$, which is the main result of Ref.~\cite{JiePRL}. By introducing the second magnon mode $m_2$ interacting with the cavity via the state-swap interaction $H_{a m_2} \,{=}\,\,\hbar g_2 (a\, m_2^{\dag} + a^{\dag} m_2)$, the two magnon modes therefore get entangled.

In the frame rotating at the magnon drive frequency $\omega_0$, the QLEs describing the system are given by
\begin{equation}
\begin{split}
\dot{a}&= - (i \Delta_a + \kappa_a) a - i \sum_{j=1,2} g_j m_j + \sqrt{2 \kappa_a} a^{\rm in},  \\
\dot{m}_1&= - (i \Delta_1 + \kappa_1) m_1  - i G_0 m_1 q - i g_1 a  + \Omega + \!\! \sqrt{2 \kappa_1} m_1^{\rm in},  \\
\dot{m}_2&= - (i \Delta_2 + \kappa_2) m_2 - i g_2 a + \sqrt{2 \kappa_2} m_2^{\rm in},  \\
\dot{q}&= \omega_b p,   \\    
\dot{p}&= - \omega_b q - \gamma_b p - G_0 m_1^{\dag}m_1 + \xi, 
\end{split}
\end{equation}
where $\Delta_{a}=\omega_{a} - \omega_0$, $\Delta_{j}=\omega_{j}-\omega_0$ ($j\,{=}\,1,2$), $\kappa_a$, $\kappa_j$, and $\gamma_b$ are the dissipation rates of the cavity, magnon, mechanical modes, respectively, and $a^{\rm in}$, $m_j^{\rm in}$ are input noise operators affecting the cavity and magnon modes, respectively, which are zero mean and characterized by the following correlation functions~\cite{Zoller}: $\langle a^{\rm in}(t) \, a^{\rm in \dag}(t')\rangle \,\,{=}\,\, \big[ N_a(\omega_a)\, {+}\,1 \big] \,\delta(t\,{-}\,t')$, $\langle a^{\rm in \dag}(t) \, a^{\rm in}(t')\rangle \,\,{=}\,\, N_a(\omega_a) \, \delta(t\,{-}\,t')$, and $\langle m_j^{\rm in}(t) \,\, m_j^{\rm in \dag}(t')\rangle = \big[ N_j (\omega_j) \,\,{+}\,1 \big] \, \delta(t\,{-}\,t')$, $\langle m_j^{\rm in \dag}(t) \, m_j^{\rm in}(t')\rangle \,\,{=}\,\, N_j(\omega_j)\,\, \delta(t\,{-}\,t')$. The Langevin force operator $\xi$, which accounts for the Brownian motion of the mechanical mode, is autocorrelated as $\langle \xi(t)\xi(t')\,\,{+}\,\,\xi(t') \xi(t) \rangle/2 \,\,\, {\simeq} \,\,\, \gamma_b \,\, \big[ 2 N_b(\omega_b) \,\,{+}\,\, 1 \big] \,\, \delta(t\,{-}\,t')$, where we have made the Markov approximation, which is a good approximation for a mechanical oscillator of a large quality factor ${Q}_b \,\,{=}\,\, \omega_b/\gamma_b \,\, {\gg}\, 1$~\cite{Markov}, and $N_k(\omega_k) \,\,{=}\,\, \Big[ {\rm exp}\Big( \frac{\hbar \omega_k}{k_B T} \Big) {-}1 \Big]^{-1} $, $k\,{=}\,a,j,b$, are the equilibrium mean thermal photon, magnon, and phonon number, respectively, with $k_B$ the Boltzmann constant and $T$ the environmental temperature.

\section{Steady-state solutions and the results of magnon entanglement}
\label{results}

Since the magnon mode $m_1$ is strongly driven by an external microwave field, and owing to the beamsplitter interactions between the cavity and the two magnon modes, both the cavity and magnon modes are of large amplitudes, $|\langle a \rangle|, |\langle m_j \rangle| \gg 1$. This allows us to linearize the dynamics of the system around the steady-state values by writing any operator as $O=\langle O \rangle +\delta O$, ($O\, {=}\, a,m_j,q,p$), and neglecting small second-order fluctuation terms. Since we are particularly interested in the quantum correlation properties of the two magnon modes, we focus on the dynamic of the quantum fluctuations of the system. The linearized QLEs describing the fluctuations of the system quadratures $(\delta X, \delta Y, \delta x_1, \delta y_1, \delta x_2, \delta y_2, \delta q, \delta p)$, with $\delta X=(\delta a + \delta a^{\dag})/\sqrt{2}$, $\delta Y=i(\delta a^{\dag} - \delta a)/\sqrt{2}$, $\delta x_j=(\delta m_j + \delta m_j^{\dag})/\sqrt{2}$, and $\delta y_j=i(\delta m_j^{\dag} - \delta m_j)/\sqrt{2}$, can be written in the form of
\begin{equation}
\dot{u} (t) = A u(t) + n(t) ,
\end{equation}
where $u(t){=}\big[\delta X (t),\! \delta Y (t),\! \delta x_1 (t),\! \delta y_1 (t),\! \delta x_2 (t),\! \delta y_2 (t),\! \delta q (t),\! \delta p (t) \big]^T$, $n (t) {=} \big[ \!\sqrt{2\kappa_a} X^{\rm in} (t), \! \sqrt{2\kappa_a} Y^{\rm in} (t), \! \sqrt{2\kappa_1} x_1^{\rm in} (t), \! \sqrt{2\kappa_1} y_1^{\rm in} (t), \! \sqrt{2\kappa_2} x_2^{\rm in} (t), \\  \! \sqrt{2\kappa_2} y_2^{\rm in} (t), 0, \xi (t) \big]^T$ is the vector of input noises, and the drift matrix $A$ is given by  
\begin{equation}\label{AAA}
A =
\begin{pmatrix}
-\kappa_a  & \, \Delta_a  \, &  0 &  \, g_1 \, &  0 &  g_2  &  0  &  0   \\
-\Delta_a  & \, -\kappa_a  \, & -g_1  & \, 0 \, & -g_2  & 0  &  0  &  0   \\
0 & g_1  & -\kappa_1  & \tilde{ \Delta}_1 & 0  &  0  & -G  &  0 \\
-g_1  & 0 & -\tilde{ \Delta}_1 & -\kappa_1 & 0  &  0  &  0  &  0 \\
0 & g_2  & 0  &  0  & -\kappa_2  &  \Delta_2 &  0  &  0 \\
-g_2  & 0 & 0  &  0  &  -\Delta_2 & -\kappa_2 & 0  &  0 \\
0 &  0  &  0  &  0  &  0  & 0  &  0  &  \omega_b   \\
0 &  0  &  0  &  G  & 0  &  0  & -\omega_b & -\gamma_b   \\
\end{pmatrix} ,
\end{equation}
where $\tilde{ \Delta}_1 \,{=}\, \Delta_1 \,{+}\, G_0 \langle q \rangle$, with $\langle q \rangle \,{=}\, {-} \frac{G_0}{\omega_b} |\langle m_1 \rangle|^2 $, is the effective detuning for the first magnon mode, which includes the frequency shift due to the magnon-phonon interaction, and $G \, {=}\, i \sqrt{2} G_0 \langle m_1 \rangle$ is the effective magnomechanical coupling rate, where $\langle m_1 \rangle \,{\simeq}\, {-}\, \big( g_1 \langle a \rangle + i \Omega \big)/\tilde{ \Delta}_1$, which implies that the coupling can be significantly enhanced by a strong driving field. This is a result of the nonlinear term $G_0 m_1^{\dag} m_1 q$ in the Hamiltonian Eq.~\eqref{Hamil}.  The average $\langle a \rangle$ is given by  
\begin{equation}\label{Ava}
\langle a \rangle \simeq  \frac{ i g_1 \Delta_2  \Omega} {\Delta_a \tilde{ \Delta}_1 \Delta_2  - g_1^2 \Delta_2 - g_2^2 \tilde{ \Delta}_1},
\end{equation}
and therefore we can also obtain $\langle m_2 \rangle$ via $\langle m_2 \rangle \simeq -g_2 \langle a \rangle /\Delta_2$. The magnomechanical coupling $G$ is a key parameter and we thus provide its specific expression, i.e.,
 \begin{equation}\label{Gexp}
G \simeq  \frac{ \sqrt{2} G_0 \Omega (g_2^2  - \Delta_2 \Delta_a)} { g_2^2 \tilde{ \Delta}_1 + \Delta_2 (g_1^2  - \tilde{ \Delta}_1 \Delta_a) }.
\end{equation}
The above expressions of $\langle a \rangle $, $\langle m_j \rangle$, and $G$ are achieved under the condition that $|\Delta_a|,  |\tilde{ \Delta}_1|,  |\Delta_2|  \gg  \kappa_a, \kappa_1, \kappa_2 $, and in this instance, $\langle a \rangle $ and $\langle m_j \rangle$ are pure imaginary numbers. The drift matrix $A$ of Eq.~\eqref{AAA} is given under this condition. In fact, we will show later that $|\Delta_a|, \, |\tilde{ \Delta}_1|, \, |\Delta_2| \simeq \omega_b  \gg  \kappa_a, \kappa_1, \kappa_2$ [see Fig.~\ref{fig1} (c)] are optimal for generating the entanglement between the two magnon modes.

Owing to the linearized dynamics and the fact that all noises are Gaussian, the dynamical map of the system preserves the Gaussian nature of any input state. The steady state of the quantum fluctuations of the system is thus a continuous variable four-mode Gaussian state, which is completely characterized by an $8\times8$ covariance matrix (CM) $\VV$, which is defined as $\VV_{ij}=\frac{1}{2}\langle u_i(t) u_j(t') + u_j(t') u_i(t)   \rangle$ ($i,j=1,2,...,8$). The stationary CM $\VV$ can be straightforwardly obtained by solving the Lyapunov equation~\cite{DV07,Hahn}
\begin{equation}\label{Lyap}
A \VV+\VV A^T = -D,
\end{equation}
where $D={\rm diag} \big[ \kappa_a (2N_a+1), \kappa_a (2N_a+1), \kappa_1 (2N_1+1),  \kappa_1 (2N_1+1), \kappa_2 (2N_2+1),  \kappa_2 (2N_2+1), 0,  \gamma_b (2N_b +1 ) \big]$ is the diffusion matrix, which is defined by $\langle  n_i(t) n_j(t') +n_j(t') n_i(t) \rangle/2 = D_{ij} \delta (t-t')$. We adopt the logarithmic negativity~\cite{LogNeg} to quantify the magnon entanglement, which is a full entanglement monotone under local operations and classical communication~\cite{Plenio} and sets an upper bound for the distillable entanglement~\cite{LogNeg}. The logarithmic negativity is defined as~\cite{Adesso}
\begin{equation}
E_N \equiv \max[0, \, -\ln2\tilde\nu_-],
\end{equation}
where $\tilde\nu_-\,\,{=}\,\min{\rm eig}|i\Omega_2\tilde{\VV}_4|$ (with the symplectic matrix $\Omega_2=\oplus^2_{j=1} \! i\sigma_y$ and the $y$-Pauli matrix $\sigma_y$) is the minimum symplectic eigenvalue of the CM $\tilde{\VV}_4={\cal P}_{1|2}{\VV_4}{\cal P}_{1|2}$, where $\VV_4$ is the $4\times 4$ CM of the two magnon modes, obtained by removing in $\VV$ the rows and columns of the cavity and mechanical modes, and ${\cal P}_{1|2}={\rm diag}(1,-1,1,1)$ is the matrix that realizes partial transposition at the level of CMs~\cite{Simon}.

\begin{figure}[b]
\hskip-0.18cm\includegraphics[width=\linewidth]{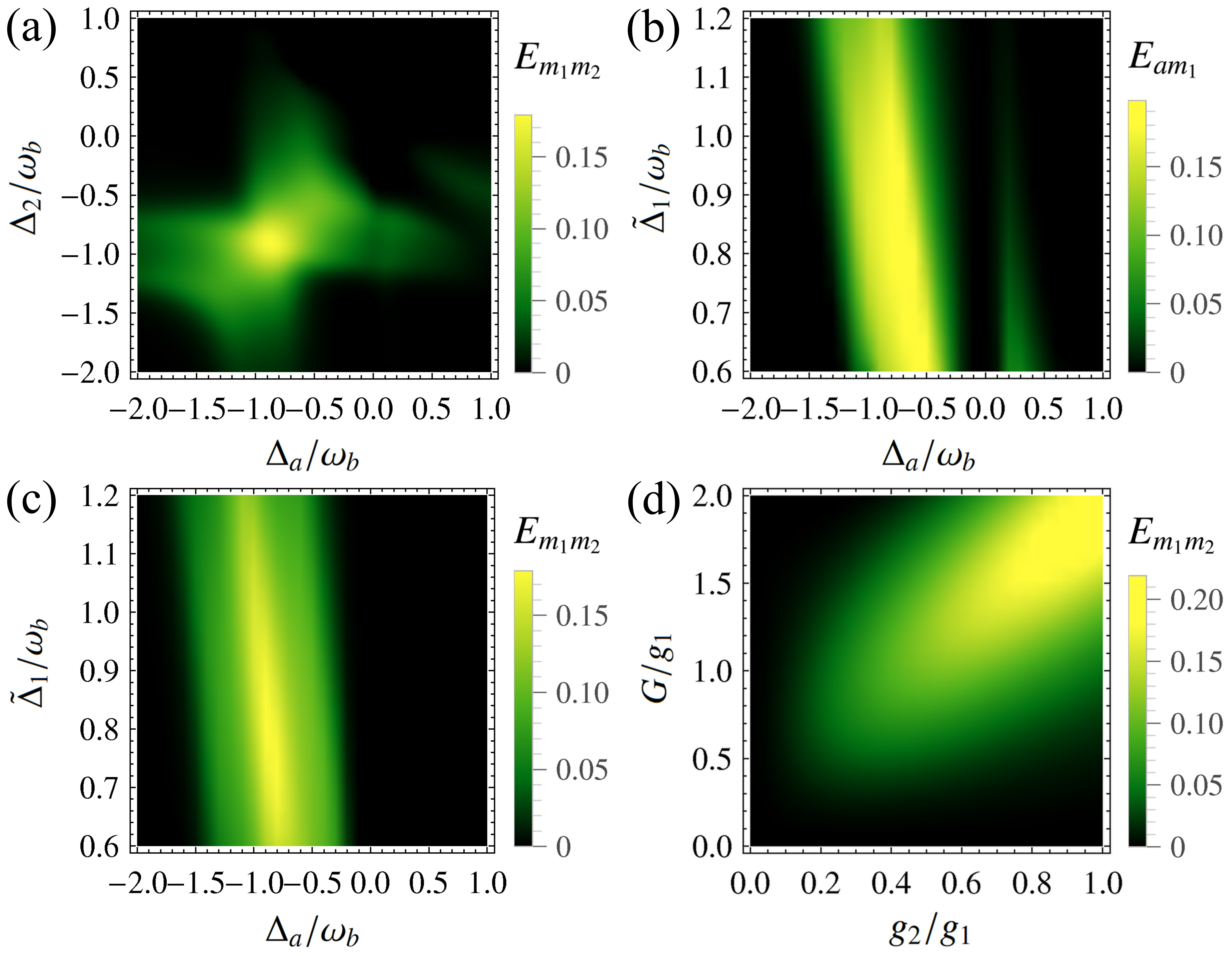} 
\caption{Density plot of the entanglement $E_{m_1m_2}$ between two magnon modes vs (a) $\Delta_a$ and $\Delta_2$, (c) $\Delta_a$ and $\tilde{ \Delta}_1$, and (d) the ratios of $g_{2}/g_{1}$ and $G/g_{1}$ ($g_{1}$ is fixed). (b) The entanglement $E_{am_1}$ between cavity and magnon mode $m_1$ vs $\Delta_a$ and $\tilde{ \Delta}_1$ with $g_2\,{=}\,0$. We take $\tilde{ \Delta}_1\,{=}\,0.85\omega_b$ in (a) and (d), $g_2/2\pi\,{=}\,2.6$ MHz in (a) and (c), $\Delta_a\,{=}\,{-}\,0.9 \omega_b$ in (d), and $\Delta_2\,{=}\,\Delta_a$ in (c) and (d). See text for details of the other parameters.}
\label{fig2}
\end{figure}

\begin{figure}[t]
\hskip-0.15cm\includegraphics[width=0.97\linewidth]{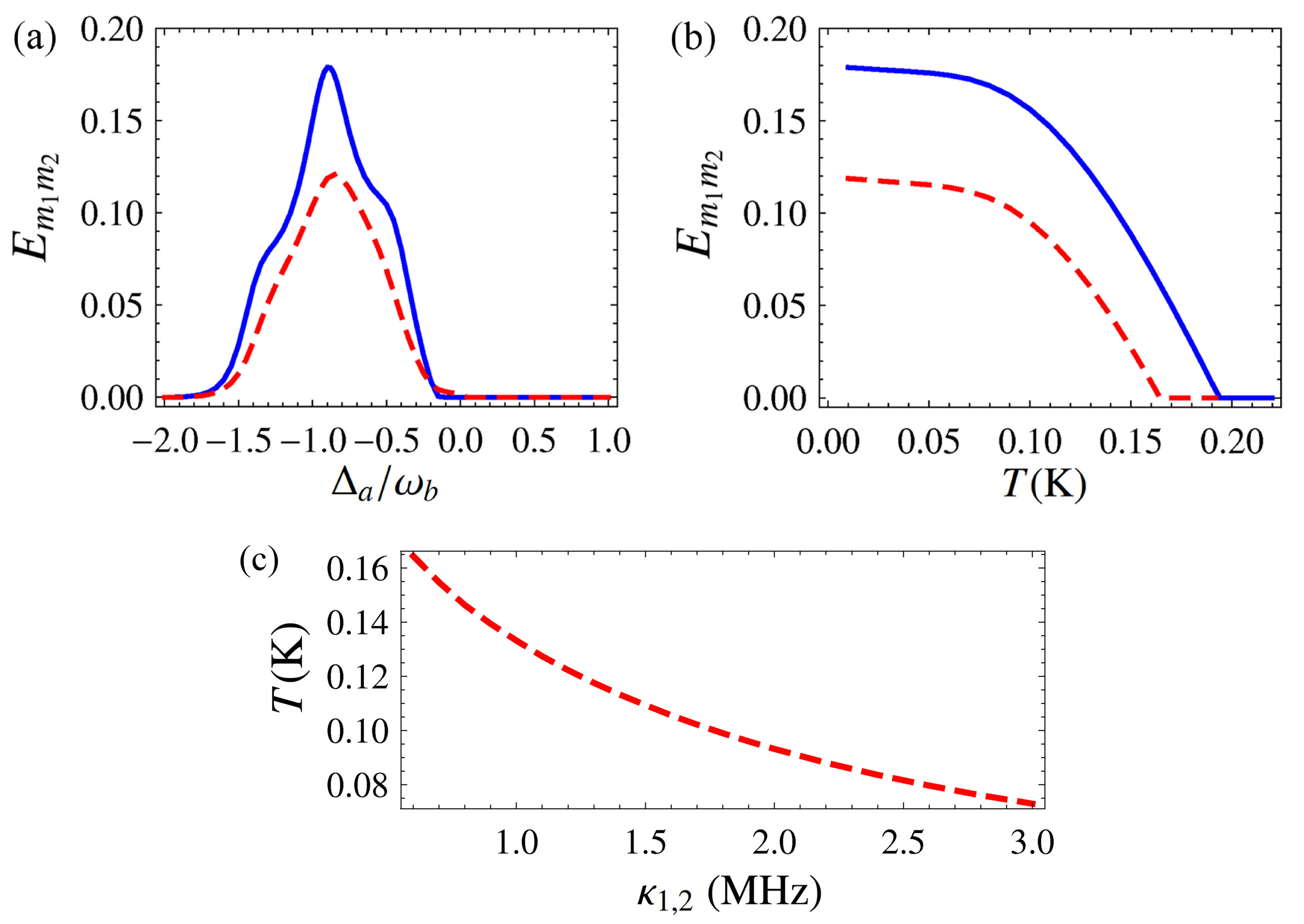}
\caption{Magnon entanglement $E_{m_1m_2}$ vs (a) $\Delta_a$ at 10 mK, and (b) temperature for the two cases of $\kappa_{a}/2\pi=\kappa_{1(2)}/2\pi=1$ MHz (solid lines) and $\kappa_{a}/2\pi=5\kappa_{1(2)}/2\pi= 3$ MHz (dashed lines). (c) Critical temperature (below which $E_{m_1m_2}\,{>}\,0$) vs $\kappa_{1(2)}/2\pi$, with always $\kappa_{a}=5\kappa_{1(2)}$. We take an optimal detuning $\Delta_a=-0.9 \omega_b$ in (b) and (c). The other parameters are as in Fig.~\ref{fig2} (a) with $\Delta_2=\Delta_a$.}
\label{fig3}
\end{figure}

Figure~\ref{fig2} shows the entanglement of the two magnon modes versus some key parameters of the system, which is in the steady state guaranteed by the negative eigenvalues (real parts) of the drift matrix $A$. We have employed experimentally feasible parameters~\cite{Tang16}: $\omega_a/2\pi \,\,{=}\,10$ GHz, $\omega_b/2\pi \,\,{=}\,10$ MHz, $\gamma_b/2\pi \,\,{=}\,\,10^2$ Hz, $\kappa_a/2\pi\,\,{=}\,\,1$ MHz, $\kappa_{1(2)}\,\,{=}\,\,\kappa_a$, $g_{1}/2\pi \,\,{=}\,\, 3.2$ MHz, $G/2\pi \,\,{=}\,\,4.8$ MHz, and at low temperature $T\,{=}\,10$ mK. We take an optimal detuning $\tilde{ \Delta}_1 \,\, {\simeq}\,\, \omega_b$~\cite{JiePRL} for the first magnon mode, which is responsible for significantly cooling the mechanical mode as quantum entanglement survives with only small thermal phonon occupancy. Combining with a {\it strong} magnomechanical coupling $G$~\cite{Note0},  magnon-phonon entanglement ($E_{m_1b}$) is created and then partially transferred to the cavity-magnon ($m_1$) subsystem, i.e., the cavity and magnon mode $m_1$ get entangled, $E_{am_1}\,{>}\,0$~\cite{JiePRL}. This effect is prominent when the cavity detuning $\Delta_a \,\,{\simeq}\, {-}\,\omega_b$. By coupling the second magnon mode ($m_2$) to the cavity, and using their state-swap interaction, the two magnon modes are expected to be entangled. This is confirmed by Fig.~\ref{fig2} (a) and it manifests that the optimal situation is that the magnon mode is resonant with the cavity, $\Delta_2 \,\, {\simeq}\,\, \Delta_a \,\,{\simeq}\,\, {-}\omega_b$ [see Fig.~\ref{fig1} (c)]. This is also the case for generating squeezed states of magnons by driving the cavity with a squeezed microwave field~\cite{JieRapid}. Figure~\ref{fig2} (b) and (c) denote the cavity-magnon ($m_1$) entanglement $E_{am_1}$ with $g_2\,{=}\,0$ and the magnon-magnon entanglement $E_{m_1m_2}$ with $g_2\,{\neq}\,0$, respectively. The similar patterns of Fig.~\ref{fig2} (b) and (c) show more clearly the magnon entanglement $E_{m_1m_2}$ is transferred from the cavity-magnon ($m_1$) entanglement $E_{am_1}$ due to the state-swap interaction between the cavity and the magnon mode $m_2$. We take $g_2/2\pi=2.6$ MHz in Fig.~\ref{fig2} (a) and (c), and therefore we have $g_{1}^2, g_{2}^2  \ll |\tilde{ \Delta}_1 \Delta_a |, |\Delta_2 \Delta_a | \simeq \omega_b^2$, which leads to a rather simple expression of the coupling $G \, {\simeq} \sqrt{2} G_0 \frac{\Omega}{\omega_b}$ [see Eq.~\eqref{Gexp}]. $G/2\pi =4.8$ MHz implies the drive magnetic field $B_0 \simeq 3.9 \times 10^{-5}$ T for $G_0/2\pi \simeq 0.3$ Hz, corresponding to the drive power $P\simeq 8.9$ mW~\cite{B0P}. A larger coupling $G$ would yield a larger entanglement. However, we take a moderate value to keep the system stable and avoid unwanted nonlinear effect (we analyse this in the next section). There are optimal couplings of $g_2$ and $G$ for fixed $g_1$, as shown in Fig.~\ref{fig2} (d). Since all bipartite entanglements of the subsystems originate from the magnon-phonon coupling, there is an interplay among the three couplings, $g_1$, $g_2$, and $G$, of the four modes of the system [see Fig.~\ref{fig1} (b)], which results in a maximum magnon entanglement. 
The magnon entanglement is robust against environmental temperature and survives up to about 200 mK, as shown in Fig.~\ref{fig3} (solid lines).

\section{Analysis of nonlinear effect and strategies for entanglement detection}
\label{Kerr}

It should be noted that the results of section~\ref{results} are valid only when the magnon excitation numbers $\langle m_j^{\dag} m_j \rangle  \ll  2Ns \,\,{=}\,\, 5N$. For what we used a 250-$\mu$m-diam YIG sphere, the number of spins $N \,\,{\simeq}\,\, 3.5 \times 10^{16}$, and the coupling strength $G/2\pi \,\,{=}\,\,4.8$ MHz corresponds to $|\langle m_1 \rangle| \,\,{\simeq}\,\, 1.1 \times 10^7$, and Rabi frequency $\Omega \,\,{\simeq}\,\, 7.1 \times 10^{14}$ Hz, such that $\langle m_1^{\dag} m_1 \rangle \,\,{\simeq}\,\, 1.3 \times 10^{14} \,\,{\ll}\,\, 5N\,{=}\,1.7 \times 10^{17}$, which is well satisfied. $\langle m_2^{\dag} m_2 \rangle \,\,{\ll} \,\, 2Ns$ is also well fulfilled due to the fact that $|\langle m_2 \rangle| \ll |\langle m_1 \rangle|$ for the parameters used in Figs.~\ref{fig2} and ~\ref{fig3}. The intense magnon drive may bring about unwanted nonlinear effects due to the Kerr nonlinear term ${\cal K} m^{\dag}m m^{\dag} m$ in the Hamiltonian~\cite{You18,You16}. The Kerr coefficient ${\cal K}$ is inversely proportional to the volume of the sphere. For a 1-mm-diam YIG sphere used in Refs.~\cite{You18,You16}, ${\cal K}/2\pi \approx 0.1$ nHz, which implies that ${\cal K}/2\pi \approx 6.4$ nHz for the spheres used in this paper. To keep the Kerr effect negligible, ${\cal K}|\langle m_1 \rangle|^3 \ll \Omega$ must hold. The parameters in Fig.~\ref{fig2} lead to ${\cal K}|\langle m_1 \rangle|^3 \simeq 5.8 \times 10^{13}$ Hz $\ll \Omega \simeq 7.1 \times 10^{14}$ Hz, which means that the nonlinear effects are negligible and our model is valid and a good approximation.

Finally, we discuss how to detect and measure the entanglement. The generated magnon entanglement can be quantified by measuring the CM of the two magnon modes, following the strategies used in Refs.~\cite{enOM,DV07}. In contrast with Ref.~\cite{JiePRL}, where the detection of a phonon state cannot be avoided, here we only need to measure the states of two magnon modes. The state of each magnon mode can be read out by coupling the magnons to an additional cavity, and by sending a weak microwave probe field and homodyning the cavity output of the probe field. This requires that the magnon dissipation rates $\kappa_{1,2}$ should be much smaller than the cavity decay rate $\kappa_a$, such that when the magnon drive is switched off and all cavity photons decay the magnon states remain almost unchanged, and then two probe fields are sent. The dashed lines in Fig.~\ref{fig3} (a) and (b) show the magnon entanglement for the case of $\kappa_{a}=5\kappa_{1(2)}$, where the entanglement is still there and survives up to about 150 mK. In Fig.~\ref{fig3} (a) and (b) we take $\kappa_{1(2)}/2\pi=0.6$ MHz, which is the lowest value demonstrated in the experiment~\cite{Tang16}. It would be useful to study the entanglement for larger magnon dissipation rates. In Fig.~\ref{fig3} (c) we plot the critical temperature versus $\kappa_{1(2)}/2\pi$ starting from 0.6 MHz to 3 MHz. The critical temperature means that above which the entanglement becomes zero due to the degradation of the thermal noise. We see that the entanglement is quite robust against the magnon dissipation rates and survives up to about 80 mK for $\kappa_{1(2)}/2\pi = 3$ MHz.

\section{conclusions}
\label{conc}

We have presented a protocol to entangle two magnon modes in a cavity magnomechanical system, where the two magnon modes in two YIG spheres couple to a microwave cavity mode, and one of them also couples to a vibrational mode of the sphere via magnetostrictive force. We have shown that with experimentally reachable parameters the two magnon modes can be prepared in a steady-state entangled state. The entanglement is achieved by exploiting the nonlinear magnetostrictive interaction and the linear cavity-magnon coupling. The magnon entangled states in massive YIG spheres represent genuinely macroscopic quantum states, and are thus useful for the study of quantum-to-classical transitions and tests of decoherence theories~\cite{Bassi}. In either the continuous spontaneous localization theory~\cite{CSL1,CSL2}, or the gravitationally induced collapse model~\cite{Diosi}, the strength of the postulated collapse noise is proportional to the size of the object (either mass or the number of particles it contains). The collapse effect becomes prominent when the size of the object is in macroscopic scale, which leads to spatial decoherence and localization of the superposition states, while it reproduces the standard results of quantum mechanics in microscopic scale. Therefore, preparing macroscopic quantum states is a key step to test those decoherence theories in macroscopic scale. Furthermore, the magnon entangled states can be applied to the quantum information processing based on magnonic systems~\cite{NakaReview}, and can also be used for creating entangled states of microwave fields, e.g., by coupling each magnon mode to a microwave cavity and utilizing their beamsplitter (state-swap) interaction.

{\it Note added}: After the completion of this work, independent proposals have been put forward, using different mechanisms, for entangling two magnon modes, either in ferrimagnetic YIG spheres~\cite{GSA1,GSA2,Jie19} or in an antiferromagnetic system~\cite{Yung}.

\section{Acknowledgments}
We thank G. S. Agarwal for continuous discussion and J. Q. You for carefully reading the article and providing constructive feedback. This work has been supported by the National Key Research and Development Program of China (Grants No. 2017YFA0304200 and No. 2017YFA0304202).


\begin{thebibliography}{99}
\bibitem{B0P}
The drive magnetic field $B_0$ is related to the power $P$ by $B_0=\frac{1}{R} \sqrt{\frac{2P\mu_0}{\pi c}}$~\cite{JiePRL}, where $R$ is the radius of the YIG sphere, $c$ is the speed of an electromagnetic wave propagating through the vacuum, and $\mu_0$ is the vacuum magnetic permeability.
\end{thebibliography}
\end{document}